\documentclass{article}
\usepackage{spconf,amsmath,graphicx,epstopdf,cite,color,mathtools,array,booktabs,multirow,float,url}

\title{Residual acoustic echo suppression based on efficient multi-task convolutional neural network}
\name{Xinquan Zhou$^{\star}$ \qquad Yanhong Leng$^{\star}$}
\address{$^{\star}$ Mutilmedia technology group, Bytedance Inc, China}
\begin{document}
%
\maketitle
\begin{abstract}
Acoustic echo degrades the user experience in voice communication systems thus needs to be suppressed completely. We propose a real-time residual acoustic echo suppression (RAES) method using an efficient convolutional neural network. The double talk detector is used as an auxiliary task to improve the performance of RAES in the context of multi-task learning. The training criterion is based on a novel loss function, which we call as the suppression loss, to balance the suppression of residual echo and the distortion of near-end signals. The experimental results show that the proposed method can efficiently suppress the residual echo under different circumstances.
\end{abstract}
\section{Introduction}
\label{sec:intro}

In voice communication systems, acoustic echo cancellation (AEC) is needed when the microphone, locating in an enclosed space with the speaker, is capturing the echo signals which is generated due to the coupling between the microphone and loudspeaker. Traditional AEC algorithm consists of two parts: adaptive linear filter (AF) \cite{deb2014technical} and nonlinear echo processor (NLP) \cite{kunshi2008_Residual}. Many challenges exist in AEC such as the nonlinearity caused by loudspeaker characteristics and it is not easy to find the nonlinear relationship between AF output and the far-end signal. In other words, NLP in AEC systems are highly likely to damage the near-end signal substantially in order to totally remove the residual acoustic echo.

In recent years, machine learning has been introduced to acoustic echo cancellation and suppression. Artificial neural network (ANN) with two hidden layers is utilized to estimate the residual echo based on the far-end signal and its nonlinear transformation signals \cite{schwarz2013_Spectral}. Training a Deep neural network (DNN) with the far-end signal and AF ouput signal can predict more accurate masks \cite{lee2015_Dnnbased,lei2019_Deep}. However, due to the lack of phase information, feeding the neural network with magnitude spectrum and estimating the output magnitude spectrum masks can hardly remain the near-end signal while removing all of the acoustic echo \cite{zhang2018_Deep}. Whereas adding more input features such phase spectrum makes the model overwhelmingly complicated to be employed in most personal terminals \cite{fazel2020_CADAEC, zhang2019_Deep}. In a recent study, phase-sensitive weight is used to revise the mask exploiting the phase relationship between AF output and near-end signal \cite{carbajal2018_Multipleinput}. 

In this paper,  we propose a new residual acoustic echo suppression (RAES) method using an efficient multi-task convolutional neural network (CNN) with far-end reference and AF output signal as inputs and phase-sensitive mask (PSM) as targets. A novel suppression loss is applied to balance the trade-off between suppressing residual echo and preserving near-end signal.
An accurate double talk detector (DTD) is essential even in a traditional AEC and the double talk state is estimated as an auxiliary task to improve the accuracy of mask prediction in our work. The experimental results prove that the proposed method is able to effectively suppress residual echo and significantly reduce the distortion of near-end signal in both simulated and real acoustic environments.

The rest of this paper is organized as follows. Section \ref{sec:aec_system} introduces the traditional AEC system. The proposed method is presented in Section \ref{sec:proposed_method} and the comparative experimental results are shown in Section \ref{sec:experiment}. Finally, Section \ref{sec:conclustion} concludes the paper.

\section{AEC framework}
\label{sec:aec_system}

\begin{figure}[htb]
	\centering
	\centerline{\includegraphics[width=7cm]{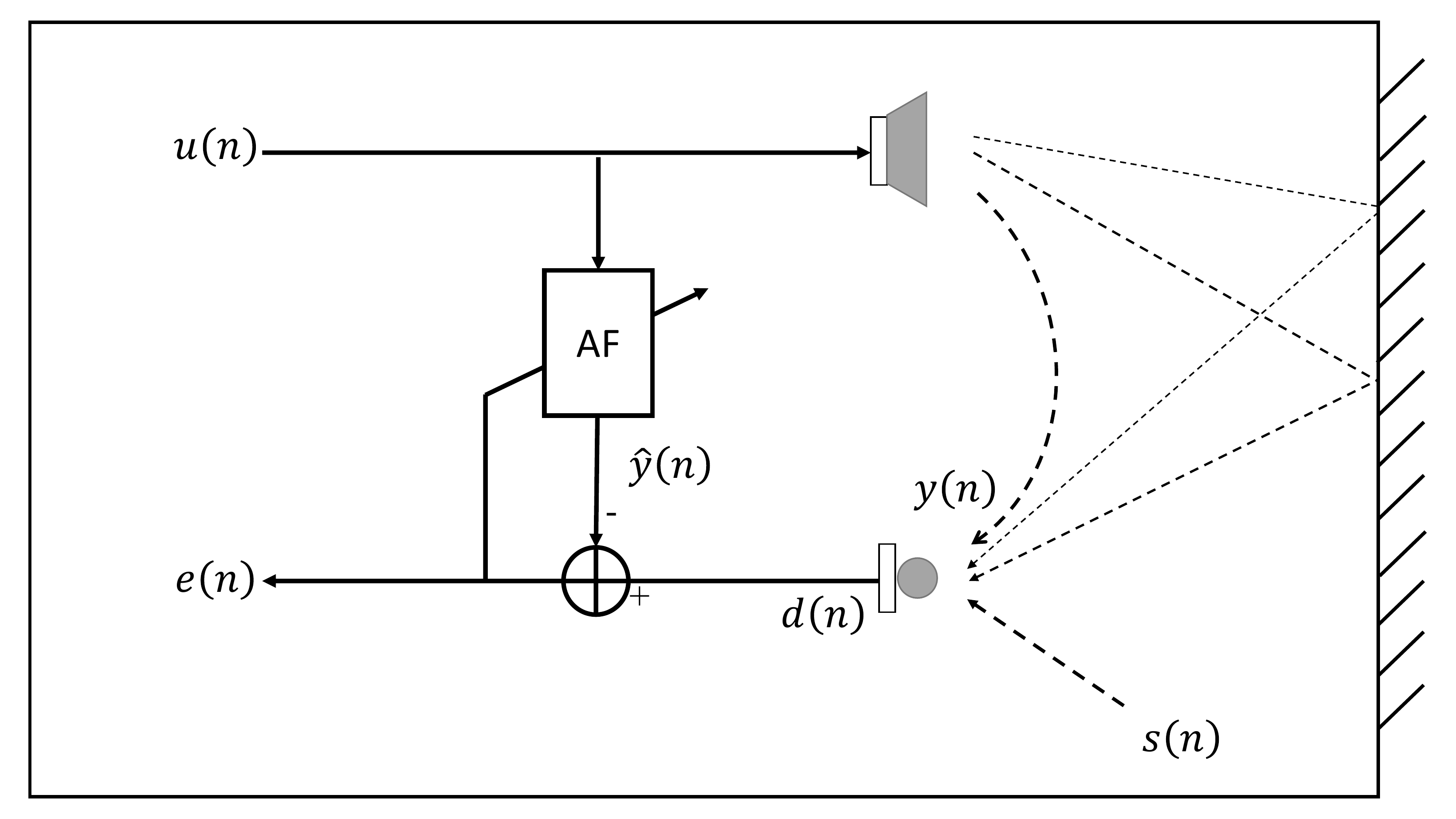}}
	\caption{Linear AEC framework.}
	\label{fig:aec_framework}
\end{figure}

In the AEC framework,  as shown in Fig. \ref{fig:aec_framework}, the signal $d(n)$ received by the microphone is composed of the near-end signal $s(n)$ and the acoustic echo $y(n)$:
\begin{align}
d(n) = s(n) + y(n). \label{eq:base}
\end{align}
The purpose of AEC is to remove the echo signal while remain the near-end signal $\hat s(n)$.

The acoustic echo $y(n)$ contains two parts: the linear echo including the direct far-end signal plus its reflected signals, and the nonlinear echo caused by the loudspeaker. The AF module adaptively estimates the linear echo $\hat y(n)$ and subtract it from the microphone signal $d(n)$ to get the output signal $e(n)$. Traditionally, the NLP calculates a suppression gain from $e(n)$ and $d(n)$ to further suppress the residual echo. However, near-end signals are highly likely to be damaged severely in the double talk segment using this kind of methods.

\section{Proposed method}
\label{sec:proposed_method}

\subsection{Feature extraction}
\label{ssec:feature}
AF module is applied to cancel a part of linear echo in the microphone signal. There are many ways to implement the linear AF algorithm. Theoretically, the proposed RAES can work with any standard AF algorithm and we use a subband normalized least-mean-square (NLMS) algorithm in this paper. 

The input feature includes the log-spectrum of the AF output error signal $e(n)$ and the far-end reference signal $u(n)$ as mentioned above. We convert $e(n)$ and $u(n)$ to the frequency domain using Short Time Fourier Transform (STFT) with a square-root Hanning window with the size of $K$. Therefore, the actual number of frequency bins is $K / 2$ with the direct current bin discarded. We concatenate $M$ frames as the input features to provide more time reference information. Another advantage of the concatenation is that it can push the network to learn the delay between echo and far-end signal.

\subsection{Network architecture}
\label{ssec:network}
\begin{figure}[htb]
	\begin{minipage}[b]{1.0\linewidth}
		\centering
		\centerline{\includegraphics[width=8cm]{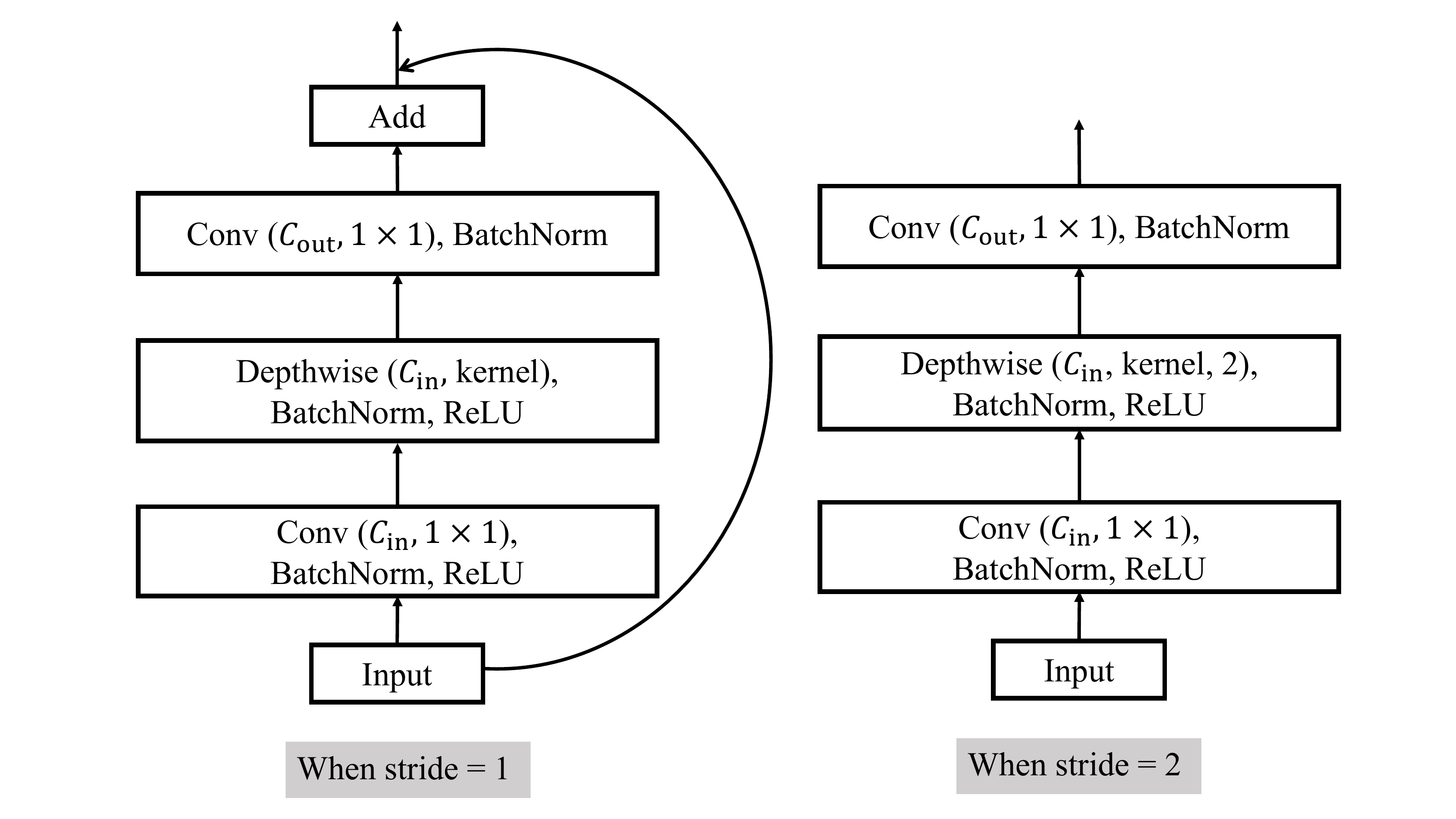}}
		\centerline{(a) Inverted Residual BottleNeck($C_{out}$, kernel, stride)}\medskip
	\end{minipage}
	
	\begin{minipage}[b]{1.0\linewidth}
		\centering
		\centerline{\includegraphics[width=8cm]{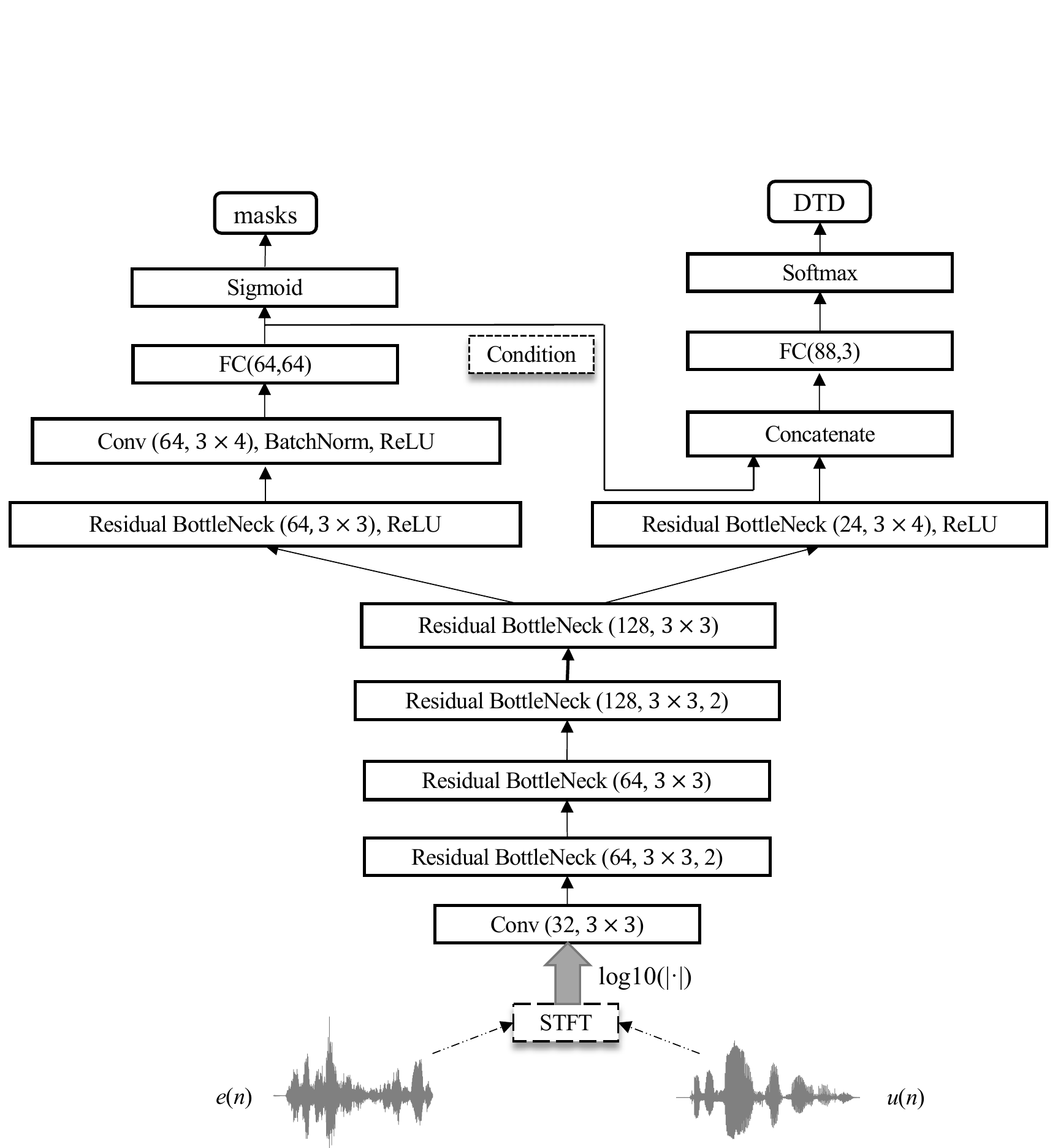}}
		\centerline{(b) Total network}\medskip
	\end{minipage}
	
	\caption{Proposed network architecture when $K = 128$.}
	\label{fig:proposed_framework}
\end{figure}

The backbone of the networks in this paper is inspired by MobileNetV2, where most of the full convolutional operation is replaced with depthwise and pointwise convolution to reduce the computational cost \cite{sandler2018mobilenetv2}. The overall network architecture is displayed in Fig.\ref{fig:proposed_framework} where the first three parameters in Conv() and Residual BottleNeck() are the number of output channel, kernel size and stride size respectively, and the default stride size is 1 if not specified. FC means full connection layer with input and output dimension. The detailed architecture of the Residual BottleNeck() is shown in Fig.\ref{fig:proposed_framework} (a),  where the residual connection fuses high-dimension and low-dimension features together. 

It is well worth mentioning that masks prediction during double talk is a challenging task. Once the features are extracted through four Residual BottleNeck blocks, we exploit a DTD prediction task in the right branch to reduce the burden on the left masks prediction branch with an conditional attention mechanism. Thus the multi-task learning can make the network focus more on the prediction of double-talk masks where masks can be set to 1 or 0 easily if the DTD task detects single talk period.

\subsection{Training targets and loss}
\label{ssec:loss}

Ideal amplitude mask (IAM) is often used as training targets in speech enhancement and residual echo suppression without considering phase information. In this paper, we use phase-sensitive mask (PSM) \cite{erdogan2015_Phasesensitive} with the expression as follows,
\begin{align}
\label{psm}
g^{\text{PSM}}(l, k) = \frac{|S(l, k)|}{|E(l, k)|} \text{cos}(\theta)
\end{align}
where $\theta = \theta^{S(l, k)} - \theta^{E(l, k)}$. $S(l, k)$ and $E(l, k)$ express near-end and AF output signal in the $l$th frame, $k$th frequency bin. PSM is truncated between 0 and 1 in the network. Then the frequency domain output of the proposed RAES $\hat S(l, k)$ in frequency bin $(l, k)$ is calculated through
\begin{align}
\hat S(l, k) = g^{\text{PSM}}(l, k)E(l, k).
\end{align}

Minimum square error (MSE) is used as the loss function in the training process. It is considered inevitable to distort near-end signal in some extent in order to remove the echo completely. As long as the estimation of the networks is not perfect, the RAES will either distort the near-end signal or remain some residual echo, or even worse, both. On the one hand, the main purpose of the AEC, intrinsically, is to remove all the echo from the microphone signal while remain the near-end signal as much as possible. Therefore, suppressing the echo is, more or less, more demanding than retaining the near-end signal quality. On the other hand, MSE loss is a symmetric metric in that the same amount of the negative and positive deviation will be counted as exactly the same loss. Therefore, using MSE directly is unable to control the trade-off between suppressing echo and preserving near-end signal. The solution in this paper is to apply a parametric leaky Rectified Linear Unit (ReLU) function to calculate a weighted mean square distance between the target and estimated mask $\Delta(l, k)$ in the frequency bin $(l, k)$ with a suppression ratio $\alpha$, 
\begin{align} \scriptsize 
\label{eq:mask_distance}
\Delta(l) = \begin{cases}
\dfrac{1}{K} \sum\limits_{k=0}^{K-1}\left[ g_t(l,k) - g_e(l,k)\right]^2 ,  & \text{if } g_t(l,k) < g_e(l,k) \\
\dfrac{1}{K} \sum\limits_{k=0}^{K-1}\left\lbrace \alpha_k \left[ g_t(l,k) - g_e(l,k) \right]\right\rbrace ^2 , & \text{else}
\end{cases}
\end{align}
where the $g_t(l,k)$ and $g_e(l,k)$ are the target and the estimated phase-sensitive mask in the frequency bin $(l, k)$ respectively, which we call it as the suppression loss. Suppression ratio $\alpha_k$ in $k$ frequency bin as a parameter is set between 0 and 1, and the smaller $\alpha_k$ is, the more severe the suppression will be. The suppression extent can be adjust in each frequency bin by setting different $\alpha_k$ value. To simplify, we just set the same $\alpha$ value in all frequency bins.

The DTD state in the $l$th frame is obtained according to the following rules:
\begin{align} \scriptsize
\text{DTD}(l) = \begin{cases} 0, & \text{if } \text{max}(|y(l, k)|) < 0.001 \text{ \& }  \text{max}(|s(l, k)|) > 0.001 \\ 1,  & \text{if } \text{max}(|s(l, k)|) < 0.001 \text{ \& }  \text{max}(|y(l, k)|) > 0.001 \\ 2, & \text{otherwise} \end{cases}
\end{align}
where the DTD states 0, 1, 2 correspond to signal near-end talk, single far-end talk and double talk respectively. Due to the imbalance between single and double talk in the dataset, focal loss\cite{lin2017focal} with focusing parameter $\gamma^{*} = 2$ is used as the loss function of DTD training task and we combine the two losses as the way in \cite{kendall2018_MultiTask} with two weights updated by the network.

\section{Experimental results}
\label{sec:experiment}

\subsection{Datasets}
In the experiments, TIMIT \cite{lamel1989speech} and THCHS30 \cite{wang2015_Thchs30} dataset are used to generate training, validating and testing datasets. In the training dataset, we randomly select 423 speakers with 4230 utterances from TIMIT and 40 speakers with 5690 utterances from THCHS30. While the validating and testing dataset includes 160 different speakers with 1600 utterances from TIMIT and 16 different speakers with 2083 utterances from THCHS30. Speakers are randomly chosen as pairs where the male-female, male-male, female-female ratio are (0.3, 0.4, 0.3) in TIMIT dataset and (0.3, 0.2, 0.5) in THCHS30 dataset respectively. One far-end signal is created by concatenating three utterances from one speaker. One utterance from another speaker is used as the near-end signal and concatenated repeatedly to the same length with the far-end signal. Moreover, considering other types of signals, especially music which greatly differs from speech in the frequency and time characteristics, often played by loudspeakers as well, we intentionally mix music signals from MUSAN \cite{snyderMUSANMusicSpeech2015} with 10\% of the far-end signals randomly. There are totally 5400 training mixtures generated where 2400 mixtures are from TIMIT and the rest are from THCHS30.

Various types of devices exhibit different nonlinear characteristics and different system intrinsic delays between the far-end and microphone signal. To simulate different devices, firstly, hard clip is applied to 70\% of the far-end signals to simulate different clips of power amplifier:
\begin{align}
\label{eq:hard_clip}
u(n)_{\text{clip}} = \begin{cases} u_{\text{max}}, & \text{if } u(n) \geq u_{\text{max}} \\ u(n), &\text{if }  -u_{\text{max}} < u(n) < u_{\text{max}}\\ -u_{\text{max}}, & \text{if }  u(n) \leq -u_{\text{max}}
\end{cases}
\end{align}
where the $u_{\text{max}}$ is randomly chosen from 0.75 to 0.99. Loudspeaker nonlinearity is simulated using the memoryless sigmoid function\cite{malikStatespaceFrequencydomainAdaptive2012}. 
\begin{align}
u_{\text{nonlinear}} = \gamma\left(\dfrac{2}{1 + \text{exp}(-ab(n))} \right)
\end{align}
where $b(n) = 1.5u_{\text{clip}}(n) - 0.3u_{\text{clip}}^2(n)$. The gain $\gamma$ is set randomly from 0.15 to 0.3. The slope $a$ is set randomly from 0.05 to 0.45 when $b(n)$ is greater than 0 and from 0.1 to 0.4 otherwise.

Then we need to generate the echo signal based on the distorted far-end signal $u_{\text{nonlinear}}$. A delay ranging from 8 ms to 40 ms is added to the distorted far-end signal to simulate the inner system delay. Both simulated and real recording room impulse response (RIR) are used to convolve with the distorted far-end signals above generating the final echo signals. The simulated RIRs are generated using the image method \cite{allen1979image} with room size $(a, b, c)$. Two typical rooms with size [6.5 m,  4.1 m,  2.95 m] and [4.2 m, 3.83 m, 2.75 m] are used in this paper. The reverberation time ranges among [0.3 s, 0.4 s, 0.5 s, 0.6 s] with sample length [2048, 2048, 4096, 4096] respectively. We generate 4 different microphone positions in each room and 5 different speaker positions around each microphone. The real recording room impulse responses are selected from AIR\cite{jeubBinauralRoomImpulse2009}, BUT\cite{szokeBuildingEvaluationReal2019} and MARDY\cite{wen2006_Evaluation} with microphone-speaker distance within 1.2 m. Ninety percent of the far-end signals are convolved with randomly selected RIRs from the simulated and real recording RIR datasets above. Half of the near-end signals in the training dataset are replaced with silent signals to generate single far-end talks. The SER during double talk period is randomly selected from -13 dB to 0 dB. The same procedure is implemented when generating validating datasets.

\subsection{Experimental configurations}
The window length of STFT is set to 128 with 50\% overlap. And 20 frames of far-end and AF output signals are concatenated forming the input features with the shape of $2 \times 20 \times 64$ with the direct current and the negative frequency bins discarded. Then we reshape it to $2 \times 40 \times 32$ as an individual input and the batch size is set to 1024. Adam optimizer is applied with initial learning rate 0.003. Suppression ratio $\alpha$ is set to 0.5 or 1.0. 

\subsection{Evaluation metrics}
The proposed method is evaluated in terms of perceptual evaluation of speech quality (PESQ)\cite{rix2001perceptual} and short-time objective intelligibility (STOI) \cite{taal2011algorithm} during double talk periods and echo return loss enhancement (ERLE) during far-end single talk periods. The ERLE in linear AEC framework is calculated by
\begin{align}
\text{ERLE} = 10 \text{log}10\left( \frac{\sum_{n}d^2(n)}{\sum_{n}e^2(n)}\right) 
\end{align}
and we extended it to measure the echo suppression extent in nonlinear RAES framework by replacing $e(n)$ with $\hat s(n)$.

\subsection{Performance comparisons}
\label{ssec:performance}

In most of hardware devices, the distance between the microphone and loudspeaker are relatively close resulting in low SER. We generate test mixtures with 0 dB, -5 dB and -10 dB SER and compare the proposed method with AEC3 in WebRTC\footnote{\url{https://webrtc.googlesource.com/src/}} and DNN method \cite{lee2015_Dnnbased}. DNN architecture is compose of three hidden layer with 2048 nodes in each layer and without restricted Boltzmann machines (RBM) being pre-trained to initialize the DNN parameter. For DNN method, we also concatenate 20 frames as input features. Ideal amplitude mask and MSE loss function are chosen for the DNN training with the same learning rate with RAES. 

We generate 50 pairs of TIMIT and 50 pairs of THCHS30 testing mixtures for each case. Table \ref{tab:far_speech} shows the ERLE results of different algorithms during single far-end talk scenarios. The proposed RAES method yield more than 40 dB ERLE showing the ability to suppress echo is better than AEC3 and DNN method especially when speech and music both exist in echo signals. What stands in the Table \ref{tab:double_speech} and \ref{tab:double_speech_music} is that PESQ and STOI scores of different methods. The scores of RAES outperforms AEC3 and DNN method, which indicate that RAES could preserve better speech quality and intelligibility during the double talk periods. The suppression ratio $\alpha$ can be used to adjust the suppression extent of the model and using smaller $\alpha$ will suppress harder on both echo and near-end signal. The F1 value of DTD task in training and validating process is 93.0\% and 90.3\% respectively. These results suggest that further post-processing can be done to the masks according to the reliable DTD.

\begin{table}
	\caption{Average ERLE(dB) during single far-end talk}
	\label{tab:far_speech}  
	\begin{center}
		\begin{tabular}{*{3}{c}}
			\toprule[1pt]
			& speech & speech + music \\
			\toprule[1pt]
			AF & 16.612 & 17.973 \\
			\hline
			AEC3 & 29.976 & 25.376 \\
			\hline
			AF+DNN & 27.832 & 31.492 \\
			\hline
			AF+RAES($\alpha=0.5$) & \textbf{40.786} & \textbf{43.144} \\
			\hline
			AF+RAES($\alpha=1.0$) & 35.597 & 36.454 \\
			\toprule[1pt]
		\end{tabular}
	\end{center}
\end{table}

\begin{table}
	\caption{Average PESQ and STOI during double talk (speech)}
	\label{tab:double_speech}  
	\begin{center}
		\begin{tabular}{*{5}{c}}
			\toprule[1pt]
			& SER & 0dB & -5 dB & -10 dB \\
			\toprule[1pt]
			\multirow{5}*{PESQ} & Origin & 1.908 & 1.519 & 1.248 \\
			\cline{2-5}
			~ & AEC3 & 1.610 & 1.520 & 1.292 \\
			\cline{2-5}
			~ & AF+DNN & 2.666 & 2.463 & 2.094 \\
			\cline{2-5}
			~ & AF+RAES($\alpha=0.5$) & \textbf{2.816} & 2.591 & 2.163 \\
			\cline{2-5}
			~ & AF+RAES($\alpha=1.0$) & 2.809 & \textbf{2.598} & \textbf{2.200} \\
			
			\toprule[1pt]
			\multirow{5}*{STOI} & Origin & 0.728 & 0.582 & 0.485 \\
			\cline{2-5}
			~ & AEC3 & 0.623 & 0.569 & 0.494 \\
			\cline{2-5}
			~ & AF+DNN & 0.856 & 0.809 & 0.727 \\
			\cline{2-5}
			~ & AF+RAES($\alpha=0.5$) & 0.875 & 0.836 & 0.760 \\
			\cline{2-5}
			~ & AF+RAES($\alpha=1.0$) &  \textbf{0.889} & \textbf{0.851} &  \textbf{0.776} \\
			\toprule[1pt]
		\end{tabular}
	\end{center}
\end{table}

\begin{table}
	\caption{Average PESQ and STOI during double talk (speech+music)}
	\label{tab:double_speech_music}  
	\begin{center}
		\begin{tabular}{*{5}{c}}
			\toprule[1pt]
			& SER & 0dB & -5 dB & -10 dB  \\
			\toprule[1pt]
			\multirow{5}*{PESQ} & Origin & 1.874 & 1.594 & 1.322\\
			\cline{2-5}
			~ & AEC3 & 1.734 & 1.563 & 1.297 \\
			\cline{2-5}
			~ & AF+DNN & 2.729 & 2.507 & 2.141 \\
			\cline{2-5}
			~ & AF+RAES($\alpha=0.5$) & \textbf{2.864} & 2.620 & 2.223\\
			\cline{2-5}
			~ & AF+RAES($\alpha=1.0$) & 2.849 & \textbf{2.626} & \textbf{2.283}\\
			
			\toprule[1pt]
			\multirow{5}*{STOI} & Origin & 0.689 & 0.610 & 0.475 \\
			\cline{2-5}
			~ & AEC3 & 0.641 & 0.583 & 0.518 \\
			\cline{2-5}
			~ & AF+DNN & 0.848 & 0.808 & 0.733 \\
			\cline{2-5}
			~ & AF+RAES($\alpha=0.5$) & 0.872 & 0.838 & 0.771 \\
			\cline{2-5}
			~ & AF+RAES($\alpha=1.0$) & \textbf{0.882} & \textbf{0.851} & \textbf{0.788} \\
			\toprule[1pt]
		\end{tabular}
	\end{center}
	
\end{table}

The computational complexity comparison is displayed in Table \ref{tab:flops}. SSE2 optimization is on in the AEC3. We run the DNN and RAES models based on a self-developed neural network inference library.
The real-time rate (RT) of DNN and RAES is 0.89 and 0.05 respectively when processing a 60 s-long speech with a 2.5 GHz CPU, x86\_64 processor, which indicates that RAES can be easily implemented on personal platforms.

\begin{table}
	\caption{Operation complexity comparison}
	\label{tab:flops}  
	\begin{center}
		\begin{tabular}{*{4}{c}}
			\toprule[1pt]
			& Model Size & MFLOPs & RT \\ 
			\toprule[1pt]
			AEC3 & - & - & 0.01 \\
			\hline
			DNN & 53.2 M & 13.8 &  0.89 \\
			\hline
			RAES & 1.2 M & 6.9 & 0.05 \\
			\toprule[1pt]
		\end{tabular}
	\end{center}
\end{table}

\section{Conclusions}
\label{sec:conclustion}
An efficient and effective multi-task residual acoustic echo suppression method is proposed. We evaluated the method in different simulated and real rooms under various SER talk situations. The experimental results show that proposed RAES can achieve better echo suppression performance than traditional echo cancellation methods and fairly easy to be deployed and run in real-time on most personal devices.



\vfill\pagebreak

\bibliographystyle{IEEEbib}
\bibliography{res}

\end{document}